\documentclass[epjST]{svjour}
\usepackage{graphicx}
\usepackage{setspace}
\usepackage[dvips,usenames]{color}
\definecolor{blue1}{rgb}{0.765,0.851,0.905}
\definecolor{blue2}{rgb}{0.584,0.560,0.905}
\definecolor{blue3}{rgb}{0.298,0.063,0.529}
\usepackage{amssymb}
\usepackage{wasysym}

\newcommand{\Bt}{{\bf B}_t}
\newcommand{\E}[1]{{\mathbb{E}\left[#1\right]}}
\newcommand{\Var}[1]{{\mathrm{Var}\left[#1\right]}}
\def\eref#1{(\ref{#1})}

\begin{document}
\title{Optimal  least-squares estimators of the diffusion
  constant from a single Brownian trajectory} \subtitle{}

\author{Denis  Boyer\inst{1}  \and   David  Dean\inst{2}  \and  Carlos
  Mej\'{i}a-Monasterio\inst{3,4}\fnmsep\thanks{\email{carlos.mejia@helsinki.fi}}
  \and Gleb Oshanin\inst{5}}
\institute{Instituto de F\'{\i}sica,  Universidad Nacional Autonoma de
  Mexico, D.F.  04510, Mexico \and Universit\'e de  Bordeaux and CNRS,
  Laboratoire Ondes et Mati\`ere d'Aquitaine (LOMA), UMR 5798, F-33400
  Talence,  France \and Laboratory  of Physical  Properties, Technical
  University of Madrid, Av.  Complutense s/n, 28040 Madrid, Spain \and
  Department  of Mathematics and  Statistics, University  of Helsinki,
  P.O.   Box  68  FIN-00014  Helsinki,  Finland  \and  Laboratoire  de
  Physique Th\'eorique  de la  Mati\`ere Condens\'ee (UMR  CNRS 7600),
  Universit\'e  Pierre et  Marie  Curie/CNRS, 4  place Jussieu,  75252
  Paris Cedex 5 France}
\abstract{ Modern developments in  microscopy and image processing are
  revolutionising  areas   of  physics,  chemistry,   and  biology  as
  nanoscale  objects  can  be  tracked  with  unprecedented  accuracy.
  However,  the price  paid for  having  a direct  visualisation of  a
  single particle trajectory with high temporal and spatial resolution
  is  a  consequent  lack  of  statistics. This  naturally  calls  for
  reliable  analytical  tools which  will  allow  one  to extract  the
  properties  specific to a  statistical ensemble  from just  a single
  trajectory.   In  this  presentation  we  briefly  survey  different
  analytical methods currently used  to determine the ensemble average
  diffusion  coefficient  from single  particle  data  and then  focus
  specifically  on  weighted  least-squares estimators,  seeking  such
  weight  functions for  which  such estimators  will possess  ergodic
  properties. Finally, we address  the question of the effects of
  disorder on such estimators.  }

\maketitle

\section{Introduction}
\label{intro}

Single  particle tracking  (SPT) can  be  traced back  to the  classic
studies of Jean Baptiste Perrin on Brownian motion \cite{perrin}. With
the advent of modern experimental techniques, recent years witnessed 
an explosion  of
different  approaches aiming at probing physical and biological
processes at the level of a single molecule.  A  SPT  experiment  uses
computer-enhanced video microscopy to  generate the time series of the
position $r_1, \ r_2,\ \ldots , \ r_N$ at times $t_1, \ t_2,\ \ldots ,
\ t_N$, of an individual  particle trajectory in a medium, (see, e.g.,
\cite{bra,saxton}).  Properly interpreted,  the information drawn from
a single,  or a finite  number of trajectories, provides  insight into
the mechanisms  and forces that drive  or constrain the  motion of the
particle.  Nowadays,  single particle tracking is  extensively used to
characterise the  microscopic rheological properties  of complex media
\cite{mason}, and  to probe the  active motion of  biomolecular motors
\cite{greenleaf}. In biological cells  and complex fluids, SPT methods
have  become  instrumental  in  demonstrating deviations  from  normal
Brownian   motion   of   passively   moving  particles   (see,   e.g.,
\cite{weber,bronstein,seisenberger,weigel,golding}).

The reliability  of the information drawn from  SPT analysis, obtained
at high temporal and spatial  resolution but at expense of statistical
sample size is not  always clear.  Time averaged quantities associated
with   a    given   trajectory   are   usually    subject   to   large
trajectory-to-trajectory fluctuations  even for simple  diffusion. For
proteins  transported in the  cell cytoplasm  or molecules  in aqueous
environments  in general,  Brownian diffusion  is the  basic transport
mechanism   \cite{saxton2}.    Notwithstanding   that   diffusion   is
ubiquitous in nature, recent  studies have suggested that transport on
cell's  membrane   \cite{pederson},  as  well  as   in  the  cytoplasm
\cite{golding},  may not  be limited  to pure  diffusion,  although the
microscopic origin of anomalous diffusion remains unclear.  For a wide
class  of anomalous  diffusions, described  by  continuous-time random
walks, time-averages  of certain particle's observables  are, by their
very nature, themselves random  variables distinct from their ensemble
averages \cite{rebenshtok,ralf}.  For example, the square displacement
time-averaged  along  a given  trajectory  differs  from the  ensemble
averaged   mean   squared   displacement\cite{ralf,he,lubelski}.    By
analyzing  time-averaged  displacements  of  a  particular  trajectory
realization,  subdiffusive motion can  actually look  normal, although
with strongly differing diffusion  coefficients from one trajectory to
another \cite{ralf,he,lubelski} and showing substantial ageing effects
\cite{barkai}.   Conflicting  results  in  the identification  of  the
underlying  transport   mechanisms  and  their   characterization  has
generated in recent years a debate on the most appropriate methodology
for the determination of the diffusion coefficient from SPT data.

Even though  standard Brownian motion  is much better  understood than
anomalous  diffusion, the  analysis of  its trajectories  is  far from
being   as  straightforward   as  one   might  think,   and   all  the
above-mentioned  troublesome  problems   persist.   For  instance,  in
bounded       systems,       substantial       manifestations       of
trajectory-to-trajectory  fluctuations  in   the  first  passage  time
phenomena have been recently revealed \cite{carlos,thiago}.

Standard fitting procedures applied to long but finite $d$-dimensional
Brownian trajectories unavoidably  lead to fluctuating estimates $D_f$
of the diffusion  coefficient, which might be very  different from the
true ensemble average value $D$,
\begin{equation}
\label{msd}
D = \frac{\E{\Bt^2}}{2 d t} \,.
\end{equation}
Using different fitting procedures,  variations by orders of magnitude
have been  observed in SPT  measurements of the  diffusion coefficient
for  diffusion  of the  LacI  repressor  protein  along elongated  DNA
\cite{austin}, in  the plasma membrane \cite{saxton}  or for diffusion
of  a single  protein in  the cytoplasm  and nucleoplasm  of mammalian
cells \cite{goulian}.  The dispersion of $D_f$ observed from different
single   particle  trajectories  results   from  the   rather  complex
environments in which the  measurements are performed. Each trajectory
will  have  its  own   thermal  history,  particle  interactions  with
different  impurities,  etc.  Moreover,  the broad  histograms for
observed  $D$  can  also  be  due to  blur  and  localization  errors,
intrinsic   of  any   experimental  measurements,   as   discussed  in
\cite{berglund,michalet,mb}.

The broad  dispersion of  diffusion constant estimates  extracted from
SPT  analysis   raises  several   important  questions:  Does   an  optimal
methodology able to determine  the diffusion coefficient from just one
single-particle trajectory  exist? If the  answer to this  question is
positive,  then what  is the  performance of  such methodology  to the
finite  length  and  finite  precision  of  the  measured  trajectory?
Clearly, it is highly desirable  to have a reliable estimator even for
the  hypothetical pure  cases, such  as, e.g.,  unconstrained standard
Brownian motion. Such an estimator must possess an ergodic property so
that its most  probable value should converge to  the ensemble average
  and   the  variance  should   vanish  as  the   observation  time
increases.

This  is  often not  the  case and  moreover,  ergodicity  of a  given
estimator is not known \textit{a priori} and has to be tested for each
particular  form of  the estimator.   Moreover, the  knowledge  of the
distribution  of such  an estimator  could provide  a useful  gauge to
disentangle  the  effects  of  the  medium complexity  as  opposed  to
variations  in  the   underlying  thermal  noise  driving  microscopic
diffusion. Recently, much effort has  been invested in the analysis of
this  challenging  problem and  several  important  results have  been
obtained  for the  estimators based  on the  time-averaged mean-square
displacement   \cite{greb1,greb2,greb3},    mean   maximal   excursion
\cite{tej},     or     the     maximum     likelihood     approximation
\cite{berglund,boyer,boyer1}.

In this  paper we first  review, in Section~\ref{sec:others},  some of
the existing  statistical methodologies used to  extract the diffusion
coefficient    of   single-particle    Brownian    trajectories.    In
Section~\ref{sec:LS}, we  focus on a family  of weighted least-squares
estimators  based  on  single-time  averages  that  we  have  recently
introduced,  and in   Section~\ref{sec:dist},  we obtain  the
distribution function  of the estimate  $D_f$ and study  its ergodic
properties.   When  the  underlying  dynamics  is  not  Brownian,  the
ergodicity properties of the  weighted least-squares estimators is not
guaranteed.   As  shown in  Section~\ref{sec:disorder},  where we
present new results on the  estimation of the diffusion coefficient of
a    Brownian    particle    moving    on    a    random    correlated
potential. Section~\ref{sec:concl} contains our final remarks.

\section{Estimating the diffusion coefficient of single trajectories}
\label{sec:others}

Consider  a  $d$-dimensional  Brownian  process  $\Bt$  with  variance
$\Var{\Bt}=2dDt$, and $D$ as defined in Eq.~\eref{msd}.
 
We start  this section by considering a  simple-minded, rough estimate
of $D_f$, defining it as the slope of the line connecting the starting
and the  end-points ${\bf B}_t$ of  a given trajectory,  namely $D_f =
{\bf  B}^2_t/2 d  t$. By  definition $\E{D_f}=D$.  A  single trajectory
diffusion  coefficient $D_f$  so defined  is a  random  variable whose
probability  density function  $P(D_f)$ is  the  so-called chi-squared
distribution with $d$ degrees of freedom, namely
\begin{equation}
\label{chi-s}
P(D_f) = \frac{1}{\Gamma(d/2)} \, \left(\frac{d}{2 D}\right)^{d/2} \, 
D_f^{d/2 - 1} \, \exp\left(- \frac{d}{2} \cdot \frac{D_f}{D} \right) \,,
\end{equation}
where $\Gamma(\cdot)$ is the Gamma-function. Clearly this distribution
diverges as $D_f  \to 0$ for $d=1$, $P(0)$  is constant for $d=2$,
and only for $d > 2$  the distribution has a bell-shaped form with the
finite most probable value $D_f^* = (1  - 2/d) D$. This means that, e.g., for
$d=3$,  the  most  likely  value  extracted  from  a  single  Brownian
trajectory is only $D_f=D/3$.

A more  refined method, known  as Least-Squares Estimator  (LSE), than
the  previous  one  consists  in  taking not  only  the  starting  and
end-points, but  the least-squares  estimate of the trajectory  in the
full time interval, say $t\in[0,T]$, namely
\begin{equation} \label{F-LSE}
F_{\mathrm{LSE}} = \int_0^T \left(\Bt^2 - l(t)\right)^2 dt \ ,
\end{equation}
where in  the simplest  case, the  dynamical law is  taken as  $l(t) =
2dD_{\mathrm{LSE}}t$.   Minimizing   \eref{F-LSE}   with  respect   to
$D_{\mathrm{LSE}}$ we obtain the LSE as
\begin{equation} \label{LSE}
D_{\mathrm{LSE}} = \frac{3}{8dT^3} \int_0^T \Bt^2 t dt \ .
\end{equation}
Another  related  method,  commonly   used  in  the  analysis  of  SPT,
experimental   data  consists   in  the   least-squares   fitting of the
time-averaged   square   displacement,   also   called   Mean   Square
Displacement  (MSD)  \cite{saxton,saxton2,goulian}.  For a  trajectory
followed through a time interval $T$, this is defined as
\begin{equation} \label{F-MSD}
F_{\mathrm{MSD}}(t) = \frac{1}{T-t}\int_0^{T-t}
\left(\mathbf{B}_{t+s} - \mathbf{B}_s\right)^2 ds \ .
\end{equation}
At    short   time    lags   $t\rightarrow0$,    the    time   average
$F_{\mathrm{MSD}}$  coincides with  the ensemble  average $\E{\Bt^2}$,
due to the ergodicity of the diffusion processes.  However, due to the
finite   length  of  experimental   trajectories,  the   MSD  analysis
\eref{F-MSD},  is performed over  a large  fraction of  intervals $t$,
even when it is clear that the small-$t$ behaviour is often restricted
only to  a very small interval  compared to the total  duration of the
trajectory  \cite{boyer}. Replacing $\Bt^2$  by the  MSD trajectory
$F_{\mathrm{MSD}}$  in   Eq.~\eref{F-LSE}  we  obtain   for  a  linear
dynamical law the MSD estimator
\begin{equation} \label{MSD}
D_{\mathrm{MSE}} = \frac{3}{8dT^3} \int_0^T F_{\mathrm{MSD}}(t) t \ dt \ .
\end{equation}
Calculating  the  MSD is  one  of the  most  popular  methods for  the
analysis of  SPT experimental data.   However, this method  presents a
number of fundamental limitations, leading to non reliable estimations
of the diffusion coefficient \cite{mb}.  Some of these limitations can
be improved by considering weighted  MSE, with a time dependent weight
that  takes  into   account  the  growth  of  the   variance  in  time
\cite{saxton2}.   Other  limitations  are  rooted to  the  fact  that,
differently than  the LSE, the  MSD estimator is a  two-time function.
This fact renders the  MSD estimator particularly fragile with respect
to the  localization errors intrinsic to  the experimental acquisition
of  the data.  This fragility  has been  only recently  recognized and
studied      by     several      authors      (see     {\em      e.g.}
\cite{berglund,michalet,mb}).  With no additional errors, the ensemble
average of  \eref{F-MSD}, {\em i.e.}, the  average of Eq.~\eref{F-MSD}
over  infinitely different  trajectories,  coincides with  $\E{\Bt^2}$
\cite{qian}.   However,  in experimental  data,  the  position of  the
particle at  any given  time is acquired  during a  finite integration
time,  yielding  a static  localization  error  $\delta$.  Since,  the
localization error involves the unknown diffusion coefficient $D$, the
ensemble average of Eq.~\eref{F-MSD} is modified as \cite{martin}
\begin{equation}
\E{F_{\mathrm{MSD}}(t)} = 2\left(dDt + \delta^2\right) \ .
\end{equation}
Additional  dynamic errors  due  to motion  blur  reduce the  previous
expression by $2D\tau/3$,  where $\tau$ is the time  interval at which
the  position   of  the  particle  is   recorded  \cite{savin}.   Full
consideration of these errors and modeling of the motion blur has been
recently    studied   in   ~\cite{berglund,michalet},    and   in
~\cite{mb} an  optimized version of the  least-squares fitting for
the MSD  has been proposed.   The authors of ~\cite{mb}  showed as
well  that  the  variance  of  this  optimized  MSE  decays  inversely
proportional to  the length  of the trajectory,  which means  that for
very long trajectories, the  estimated value of the diffusion constant
becomes trajectory independent.  Therefore, this two-time estimator is
ergodic.

A  conceptually  different fitting  procedure  has  been discussed  in
~\cite{boyer}  which  amounts   to  maximizing  the  unconditional
probability of  observing the whole trajectory  ${\bf B}(t)$, assuming
that it is drawn from a Brownian process with mean-square displacement
$2 d D t$.  This is the maximum likelihood estimate which
takes the value of $D$  that maximizes the likelihood of ${\bf B}(t)$,
defined as:
\begin{equation}
L_T = \prod_{t = 0}^T \left(4 \pi D t\right)^{-d/2} 
\exp\left({-\frac{\Bt^2}{4 D t}}\right)
\end{equation}
Minimization of the logarithm of  $L_T$ with respect to $D$ yields the
Maximum Likelihood Estimator (MLE)  for the diffusion coefficient of a
Brownian trajectory as \cite{boyer}
\begin{equation} \label{MLE}
D_{\mathrm{MLE}} =  \int_0^T \frac{\Bt^2}{t} \ dt \ .
\end{equation}
This estimator was studied in ~\cite{boyer} in one dimension, and
it  was shown  that  the MLE  is superior  to  those based  on the  LS
unweighted  minimization. As  a matter  of fact,  the  distribution of
$D_{\mathrm{MLE}}$ not only appears  narrower than the distribution of
$D_{\mathrm{LSE}}$, resulting  in a  smaller dispersion, but  also the
most probable value of the diffusion coefficient appears closer to the
ensemble average $D$ \cite{boyer}. More recently, the same conclusions
were obtained in ~\cite{boyer1} for arbitrary dimensions.

\section{Weighted   least-squares    estimators   of   the   diffusion
  coefficient}
\label{sec:LS}

In  a  recent  paper  \cite{boyer2},  we  have  studied  a  family  of
least-squares one-time estimators defined as
\begin{equation}
\label{u}
u_{\alpha} = \frac{A_{\alpha}}{T} \int^{T}_{0} \, \omega(t) \, {\bf B}^2_t \, dt,
\end{equation}
where $\omega(t)$ is the  weighting function of the form
\begin{equation}
\label{omega}
\omega(t) = \frac{1}{(t_0 + t)^{\alpha}} \,,
\end{equation}
$\alpha$ being a  tunable exponent, (positive or negative),  $t_0$ - a
lag time and $A_{\alpha}$  - the normalization constant, appropriately
chosen in such  a way that $\mathbb{E}\left\{u_{\alpha}\right\} \equiv
1$, so that
\begin{equation}
\label{aalpha}
A_{\alpha} = \frac{T}{2 d D} \left(\int^T_0 \frac{t \, dt}{(t_0 + t)^{\alpha}}\right)^{-1}\,.
\end{equation}
Such a normalization permits  a direct comparison of the effectiveness
of estimators corresponding to different values of $\alpha$.

The estimator $u_{\alpha}$ in Eq.~\eref{u} minimizes the least-squares
functionals   with  a   weighting   function  $\omega(t)   =  (t_0   +
t)^{-\alpha}$. Consider a  $d$-dimensional trajectory ${\bf B}_t$ with
$t \in [0,T]$.  To estimate  the diffusion coefficient $D_F$ from this
trajectory, one writes the least-squares functional of the form
\begin{equation}
\label{func}
F = \frac{1}{2} \int^T_0 \frac{\omega(t)}{t} \, \left({\bf B}^2_t - 2 d D_F t\right)^2 \,dt \, ,
\end{equation}
and  seeks  to  minimize  it  with  respect to  the  value  of  $D_F$,
considered   as  a  variational   parameter.   Eq.~\eref{func}   is  a
generalized   weighted  least-squares   functional,  with   a  certain
weighting  function $\omega(t)$, which  depending on  whether it  is a
decreasing or an increasing function  of $t$, will emphasize the short
time  or  the  long  time  behavior of  the  trajectory  ${\bf  B}_t$,
respectively.

After minimization of $F$ with respect to $D_F$ we find that
\begin{equation}
\frac{D_F}{D} = \left(\frac{1}{T} \int^T_0 dt \, \omega(t) \, 
{\bf B}^2_t\right)/\left(\frac{2 d D}{T} \int^T_0 dt \, t \, \omega(t) \right) \,.
\end{equation}
Finally,   identifying   the   denominator  with   the   normalization
$A_{\alpha}$ and choosing  $\omega(t) = (t_0+t)^{-\alpha}$, we recover
our  definition  \eref{u}.  Furthermore,  note  that  $\alpha  = -  1$
corresponds to the unweighted LSE  and $\alpha = 1$ corresponds to MLE
defined in  the previous section. Therefore,  our generalized estimate
\eref{u},  possesses  a nice  property  to  contain, as  particular
cases, two other commonly used estimators.

For   standard  Brownian  motion,  the   lag  time  $t_0$,
corresponding to the time at  which the measurement is started, can be
set  equal  to  zero. However,  it  is  useful  to keep  the  explicit
dependence on $t_0$ since it is equal to the resolution $\epsilon$
at which an experimental trajectory is recorded as $\epsilon = t_0/T$.
Moreover, in \cite{boyer1} it  was found that for anomalous diffusion,
or  for  Brownian  motion  in  presence  of  disorder  $t_0$  plays  a
significant role.

In  ~\cite{boyer2},  we  have  studied the  family  of  estimators
\eref{u},  and  determined  a  unique  value  of  $\alpha$  for  which
$u_{\alpha}$  is  ergodic, so  that  the  single trajectory  diffusion
coefficient  $D_F \to  D$ as  $\epsilon =  t_0/T \to  0$. In  the next
section  we   briefly  sketch   the  derivation  of   the  probability
distribution function $P(u_\alpha)$ and discuss its properties.

\section{Distribution of the weighted estimators}
\label{sec:dist}

The  fundamental  characteristic property  to  derive the  probability
distribution function $P(u_\alpha)$  is the moment-generating function
$\Phi(\sigma)$ of the random variable in Eq.~\eref{u} defined as
\begin{equation}
\label{laplace}
\Phi(\sigma) = \E{\exp\left( - \sigma u_{\alpha}\right)} \,,
\end{equation}
where $\sigma$ is positive definite parameter, $0 \leq \sigma < \infty$.

Using  the  fact  that   a  $d$-dimensional  Brownian  motion  can  be
decomposed into  a product of its $d$  one-dimensional components, the
generating function can be written as
\begin{equation}
\label{G}
\Phi(\sigma) =  G(\sigma)^d = \left(\E{\exp\left( - \frac{\sigma A_{\alpha}}{T}
      \int^{T}_{0} \omega(\tau) \, B_{\tau}^2(i) \, d\tau \right)}\right)^d.
\end{equation}

Following   \cite{boyer,boyer1},   we   introduce  an   auxiliary
functional
\begin{equation}
\label{fc}
\Psi(x,t) = \mathbb{E}^x_t\left\{ \exp\left( - \frac{\sigma A_{\alpha}}{T}
\int^{T}_{t} \, \omega(\tau) \, B_{\tau}^2  \, d\tau   \right)\right\} \ ,
\end{equation}
where the expectation is for a Brownian motion starting at $x$ at time
$t$.  Clearly, $G(\sigma) =  \Psi(0,0)$. This functional satisfies the
Feynman-Kac type formula
\begin{equation}
\Psi(x,t) = \mathbb{E}_{dB}\left\{ \Psi(x + dB_t,t + dt) 
\left(1 - \frac{\sigma A_{\alpha} \omega(t)}{T} x^2 dt\right)\right\} \, ,
\end{equation}
where  $dB_t$  is  an   infinitesimal  Brownian  increment  such  that
$\mathbb{E}_{dB}\{dB_t\}  = 0$  and $\mathbb{E}_{dB}\{dB^2_t\}  =  2 D
dt$,  and  $\mathbb{E}_{dB}$ denotes  averaging  with  respect to  the
increment  $dB_t$. Furthermore, expanding  the right-hand-side  of the
latter equation  to second order in  $dB_t$, linear order  in $dt$ and
performing averaging, we find that $\Psi(x,t)$ obeys the Schr\"odinger
like  equation  with a  harmonic,  time-dependent potential
\begin{equation}
\label{17}
\frac{\partial \Psi(x,t)}{\partial t} = - D \frac{\partial^2 \Psi(x,t)}
{\partial x^2} + \frac{\sigma A_{\alpha} \omega(t)}{T} x^2 \Psi(x,t) \, ,
\end{equation}
subject  to  boundary condition  $\Psi(x,T)  =  1$  for any  $x$.

The  solution of Eq.~\eref{17},  and thus  of the  generating function
\eref{G}, was explicitly obtained  in \cite{boyer2} for $\alpha=2$ and
$\alpha\ne2$.  In  the latter case, the  moment-generating function,
to leading order in $\epsilon=t_0/T$, is gyben by:
\begin{eqnarray}
\Phi(\sigma) & = & \left[\Gamma\left(\nu\right) \ \left(\frac{\sigma}{\chi_1}
\right)^{\frac{1-\nu}{2}} \ \! {\rm I}_{\nu - 1}\left(2 \sqrt{\frac{\sigma}{\chi_1}}
\right)\right]^{-d/2} \,,  \quad \textrm{for} \ \ \alpha < 2 \ ,\label{alpha<2}\\
\Phi(\sigma) & = & \left[\Gamma\left(1-\nu\right) \left(\frac{\sigma}{\chi_2}
\right)^{\frac{\nu}{2}} {\rm I}_{-\nu}\left(2 \sqrt{\frac{\sigma}{\chi_2}}
\right)\right]^{-d/2} \,, \quad  \textrm{for} \ \ \alpha > 2  \ , \label{alpha>2}
\end{eqnarray}
where $\nu  = 1/(2 - \mu)$, $I_{\mu}(z)$  is the modified Bessel
function \cite{abramowitz} and
\begin{equation}
\label{chi}
\chi_1 = \frac{d (2 - \alpha)}{2} \, \,\,\,{\rm and} \,\,\, \chi_2 =
 \frac{d (\alpha - 2)}{2 (\alpha - 1)} \,.
\end{equation}

For   the  particular   case   of  $\alpha=2$,   we   find  that   the
moment-generating   function  is   given   for  arbitrary   $\epsilon$
explicitly by
\begin{eqnarray}
\label{alpha=2}
\Phi(\sigma)  &=& \Big[\frac{(\delta + 1)}{2 \delta \epsilon^{(\delta-1)/2}} \Big(\Big( 1 + \frac{\delta -1}{\delta+1} \epsilon^{\delta}\Big) \, \cosh\left(\sqrt{2 a \xi \sigma}\right)  \nonumber\\ &+& \frac{\delta - 1}{2 \sqrt{2 a \xi \sigma}} \left(1 + \epsilon^{\delta}\right) \, \sinh\left(\sqrt{2 a \xi \sigma}\right)\Big) \Big]^{-d/2} \,.
\end{eqnarray}

\begin{figure}[!t]
  \centerline{\includegraphics*[width=0.85\textwidth]{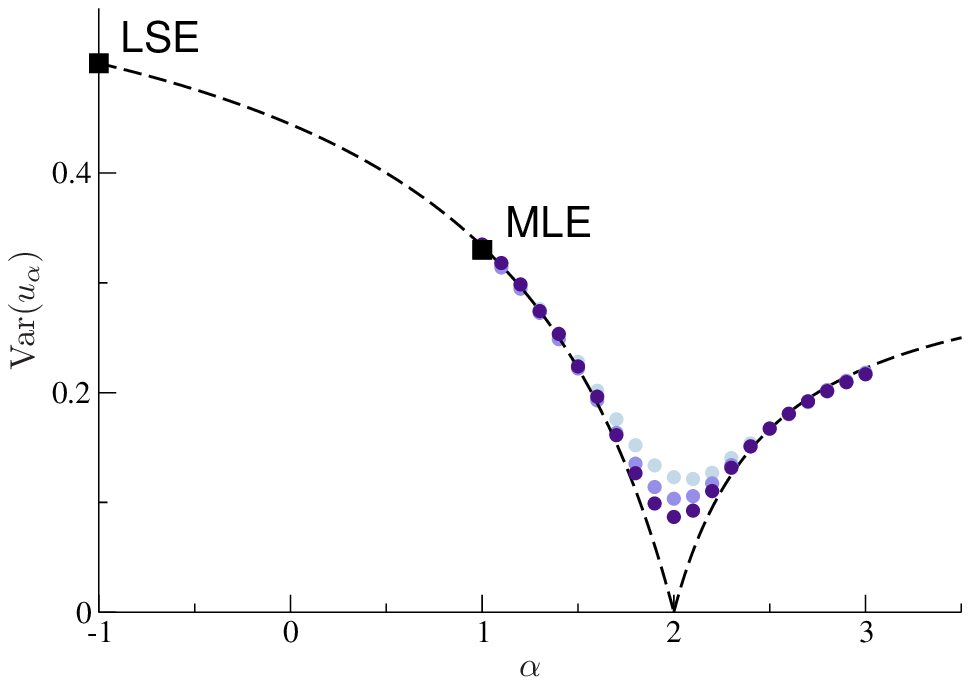}}
  \caption{Variance of the  distribution $P(u_{\alpha})$ for different
    values   of   $\alpha$.    The   dashed   curves   correspond   to
    Eq.~(\ref{varvar}).  The symbols correspond to the values obtained
    from numerical simulations  of 3D random walks for  (from light to
    dark)  $\epsilon =  5\times10^{-5}$  (${\color{blue1} \newmoon}$),
    $5\times10^{-6}$ (${\color{blue2} \newmoon}$) and $5\times10^{-7}$
    (${\color{blue3} \newmoon}$). The  solid squares correspond to the
    values  for  $\Var{u_{-1}}$   (LSE)  and  $\Var{u_{1}}$  (MLE)  as
    indicated by the labels.}
\label{fig1}
\end{figure}

First  we  focus on  the  variance  of  the estimator  given in \eref{u}.   The
variance  ${\rm Var}(u_{\alpha})$  is  obtained by
differentiating  Eqs.~(\ref{alpha<2})  or  (\ref{alpha>2}) twice  with
respect to $\sigma$ and setting $\sigma$ equal to zero.  For arbitrary
$\alpha \neq  2$ the  variance to  leading order in  $\epsilon$ is
then given explicitly by
\begin{equation}
  \label{varvar}
  \Var{u_{\alpha}} = \frac{2}{d}\left\{
    \begin{array}{ll}
      (2 - \alpha)/(3 - \alpha), & \alpha < 2 \ ,\\
      \\
      (\alpha - 2)/(2 \alpha - 3), &\alpha > 2 \ .
    \end{array}\right.
\end{equation}
The result in  the latter equation is depicted  in Fig.~\ref{fig1} and
shows that, strikingly, the variance  can be made arbitrarily small in
the leading in $\epsilon$ order by taking $\alpha$ gradually closer to
$2$, either from above or from below. The slopes at $\alpha = 2^+$ and
$\alpha =  2^-$ appear to be the  same, so that the  accuracy of the
estimator will be the same when  approaching $\alpha = 2$ from above or
below. Equation (\ref{varvar}),  although formally invalid for $\alpha
= 2$, also suggests that the estimator in Eq.~(\ref{u}) with $\alpha =
2$ possesses an ergodic property.

A word  of caution is  now in order. Finite-$\epsilon$  corrections to
the    result    in     Eq.~(\ref{varvar})    are    of    order    of
$\mathcal{O}(\epsilon^{2 - \alpha})$ for $1 < \alpha < 2$, which means
that this  asymptotic behavior can  be only attained  when $\epsilon
\ll \exp\left(-1/(2 - \alpha)\right)$.   In other words, in principle,
the variance can be made arbitrarily small by choosing $\alpha$ closer
to  $2$, but  only at  expense of  increasing either  the experimental
resolution  or the  observation time  $T$,  which is  clearly seen  in
Fig.~\ref{fig1}.

The solid  circles in Fig.~\ref{fig1} correspond  to numerical results
of  random walks  on  a  $3$-dimensional lattice  and  computed $P(u_\alpha)$  using
Eq.~(\ref{u})  from a  large ensemble  of trajectories,  for different
values of $\alpha$ and different resolution $\epsilon$.  For $\alpha<1.5$
or $\alpha>2.5$,  the variance computed numerically is  well described by
Eq.~(\ref{varvar})    and     is    independent    of    $\epsilon$
(Fig.~\ref{fig1}).   Near  $\alpha=2$,  corrections  due  to  the  finite
resolution  are noticeable,  but the  numerics clearly  show  that the
variance  of  the distribution  $P(u_\alpha)$  decreases as  $\epsilon
\rightarrow 0$.

We now turn our  attention to the distribution function $P(u_\alpha)$,
which   is   obtained   by   inverting  the   Laplace   transform   in
Eq.~(\ref{laplace}) with respect to the parameter $\sigma$:
\begin{equation}
\label{def:dist}
P(u_{\alpha}) = \frac{1}{2 \pi i} \int^{\gamma + i \infty}_{\gamma  - i \infty} 
d\sigma \, \exp\left(\sigma u_{\alpha}\right) \, \Phi(\sigma) \,,
\end{equation}
where $\gamma$ is a real number  chosen in such a way that the contour
path of integration is in the region of convergence of $\Phi(\sigma)$.
Since for $\alpha\ne2$ all the poles of the moment-generating function
lie  on   the  complex  plane  on  the   negative  real  $\sigma$-axis
\cite{boyer2,boyer3},  we can  set $\gamma=0$  in  \eref{def:dist} and
find, to leading order in $\epsilon$
\begin{equation}
\label{distr}
P(u_{\alpha}) = \frac{1}{\pi} \int^{\infty}_0 \frac{dz \, \cos\left(z u_{\alpha} - d
 \, \phi_{\alpha}(z)/2\right)}{\rho_{\alpha}^{d/4}(z)},
\end{equation}
where, for $\alpha < 2$,
\begin{eqnarray}
\rho_{\alpha}(z) = \Gamma^2\left(\nu\right) \left(\frac{\chi_1}{z}\right)^{\nu - 1}
 \, \left({\rm ber}_{\nu-1}^2\left(2 \, \sqrt{\frac{z}{\chi_1}}\right)+
{\rm bei}_{\nu -1}^2\left(2 \, \sqrt{\frac{z}{\chi_1}}\right)\right) \,,
\end{eqnarray}
and the phase $\phi$ is given by
\begin{equation}
\phi_{\alpha}(z) = {\rm arctg}\left({\rm ber}_{\nu - 1}\left(2 \,
\sqrt{\frac{ z}{\chi_1}}\right)/{\rm ber}_{\nu-1}\left(2 \, 
\sqrt{\frac{z}{\chi_1}}\right)\right) \,,
\end{equation}
while for $\alpha > 2$ we have
\begin{eqnarray}
\rho_{\alpha}(z) = \Gamma^2\left(1-\nu\right) \left(\frac{\chi_2}{z}\right)^{-\nu}
 \, \left({\rm ber}_{-\nu}^2\left(2 \, \sqrt{\frac{z}{\chi_2}}\right)+
{\rm bei}_{-\nu}^2\left(2 \, \sqrt{\frac{z}{\chi_2}}\right)\right) \,,
\end{eqnarray}
and
\begin{equation}
\phi_{\alpha}(z) = {\rm arctg}\left({\rm ber}_{-\nu}\left(2 \, \sqrt{\frac{z}
{\chi_2}}\right)/{\rm ber}_{-\nu}\left(2  \sqrt{\frac{z}{\chi_2}}\right)\right) \,,
\end{equation}
where  ${\rm ber}_{\mu}(x)$  and ${\rm  bei}_{\mu}(x)$ are  the Kelvin
functions \cite{abramowitz}.

\begin{figure}[!t]
  \centerline{\includegraphics*[width=0.95\textwidth]{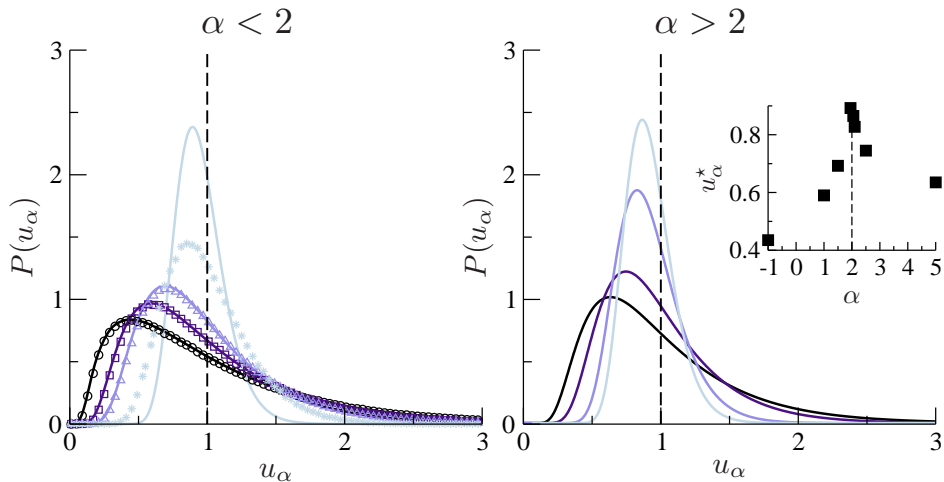}}
  \caption{The distribution $P(u_{\alpha})$ for different $\alpha \neq
    2$ in  three dimensions.   The curves from  the left to  the right
    (darker to lighter) in the left panel correspond to $\alpha = - 1$
    (LSE),  $\alpha = 1$  (MLE), $\alpha  = 3/2$  and $\alpha  = 1.95$
    (blue), and  in the  right panel  to $\alpha =  5$, $\alpha  = 3$,
    $\alpha = 2.5$ and $\alpha = 2.05$.  The symbols in the left panel
    correspond  to  numerical simulations  of  random  walks in  three
    dimensions.   In  the  inset  we  show  the  most  probable  value
    $u^\star_\alpha$ as a function of $\alpha$.}
\label{fig2}
\end{figure}

In  Figs.\ref{fig2} we plot  $P(u_{\alpha})$ in  Eq.~(\ref{distr}) for
three-dimensional systems.   Indeed, the most  probable value $u^\star
\to 1$ when $\alpha \to 2$  either from above or from below. Note that
for any  exponent $\alpha\ne2$, $u^\star$ is smaller  than the average
value  1.  Nevertheless,  already for  $\alpha =  1.95$ (or  $\alpha =
2.05$) we  get the most  probable value $u^\star \approx  0.94$, which
outperforms  the LSE ($u^\star  \approx 0.44$)  and the  MLE ($u^\star
\approx 0.6$).  For $\alpha  = 1.95$ the variance ${\rm Var}(u_{\alpha
})  \approx 0.032$,  which  is an  order  of magnitude  less than  the
variances observed  for LSE  ($= 0.5$) and  the MLE  ($\approx 0.33$).
Similarly   to  Fig.~\ref{fig1},  finite-resolution   corrections  are
negligible for  $\alpha<1.5$ and $\alpha>2.5$,  and $P(u_{\alpha})$ is
well described  by Eq.~(\ref{alpha<2}).  For  $\alpha=1.95$ and finite
resolution  $\epsilon=10^{-7}$, we obtain  a broader  distribution and
with a smaller $u^*$ than the corresponding to Eq.~(\ref{alpha<2}) for
infinite resolution.   Note, however, that the most  probable value of
$P(u_{1.95})$ that  we obtain at finite resolution  is $\approx 0.84$,
which  outperforms the  LSE and  MLE  for infinite  resolution.  As  a
matter of fact, it is evident from Fig.~\ref{fig1} that for any finite
resolution, at least $<5\times10^{-5}$, the variance of the weighted LSE
$u_\alpha$  outperforms  the  unweighted   LSE  and  MLE  at  infinite
resolution.  A  comparison between  LSE, MLE and  our weighted  LSE is
shown in Table~\ref{tab:1}.

\begin{center}
\begin{table}[!t]
  \caption{Most probable value $u^\star_\alpha$ and variance of the normalized
    estimator $u_\alpha$, for different values of $\alpha$ and $\epsilon\to0$.}
\label{tab:1}       
\begin{tabular}{lll}
\hline\noalign{\smallskip}
Estimator & $u^\star_\alpha$ & ${\rm Var}(u_{\alpha})$  \\
\noalign{\smallskip}\hline\noalign{\smallskip}
LSE $u_{-1}$ & $\approx 0.44$ &  $0.5$ \\
MLE $u_{1}$ & $\approx 0.6$ & $\approx 0.33$ \\
weighted BM $u_{1.95}$ & $\approx 0.94$ & $\approx 0.032$\\
\noalign{\smallskip}\hline
\end{tabular}
\end{table}
\end{center}

When $\alpha = 2$ and $\epsilon  = t_0/T$ small but finite we consider
a slightly more general form for $\omega(t)$:
\begin{equation}
\label{omlog}
\omega(t) = \left\{
\begin{array}{ll}
2 \xi/t_0^2, & \quad \textrm{for} \ \ t < t_0, \\
\\
1/t^2, & \quad \textrm{for} \ \  t_0 \leq t \leq T,
\end{array}\right.
\end{equation}
where  $\xi$ is a  tunable amplitude.  For such  a choice,  the moment
generating function is given explicitly by \cite{boyer2}
\begin{equation}
\label{12}
\Phi(\sigma) = \left(\frac{2 \, \delta \,
\epsilon^{(\delta-1)/2}}{\phi_{+}}\right)^{d/2} \left[1 + \frac{\phi_-}{\phi_+} \,
\epsilon^{\delta}\right]^{-d/2} \,,
\end{equation}
with
\begin{equation}
\phi_{\pm} = \left(\delta \pm 1\right) \left({\rm ch}
\left(\sqrt{2 \gamma \xi \sigma}\right) \pm \frac{\delta \mp 1}
{2 \sqrt{2 \gamma \xi \sigma}} {\rm sh}\left(\sqrt{2 \gamma \xi \sigma}\right)
\right) \,, \nonumber\\
\end{equation}
where $\delta = \sqrt{1 + 4 \gamma \sigma}$ and $\gamma = 2/d (\xi +
\ln(1/\epsilon))$. Differentiating Eq.~(\ref{12}), we find
\begin{equation}
\label{7}
{\rm Var}[u_2] = \frac{4}{3 d} \, \frac{3 \ln(1/\epsilon) - 3 (1-\epsilon) +
2 (1 - \epsilon) \xi + \xi^2}{\left(\xi + \ln(1/\epsilon)\right)^2} \, .
\end{equation}

\begin{figure}[!t]
  \centerline{\includegraphics*[width=0.85\textwidth]{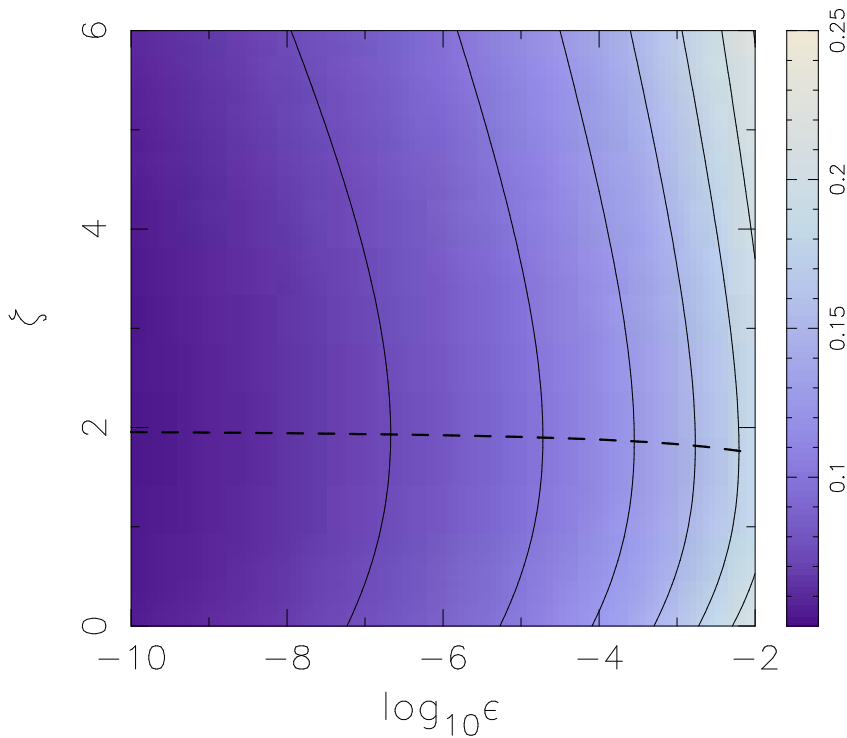}}
  \caption{Colour density  plot of ${\rm  Var}[u_2]$ as a  function of
    $\xi$ and  $\epsilon$. The dashed curve  corresponds to $\xi_{{\rm
        opt}}$ of Eq.~\eref{xi-opt}, and thus, to the optimal value of
    ${\rm Var}[u_2]$ for fixed $\epsilon$.}
\label{fig3}
\end{figure}

Now it is required to minimize \eref{7} with respect to $\xi$. We note
first   that  ${\rm   Var}(u_2)$  is   a  non-monotonic   function  of
$\xi$.  However, we  find  that the  optimal  value of  $\xi$ is,  for
arbitrary $\epsilon$
\begin{equation} \label{xi-opt}
\xi = \xi_{{\rm opt}} = \frac{(2 + \epsilon) \ln(1/\epsilon) - 3 (1 - \epsilon)}
{\ln(1/\epsilon) + \epsilon - 1} \,.
\end{equation}
This  function and  its optimal  value are  shown  in Fig.~\ref{fig3}.
Therefore, corresponding optimized variance is
\begin{equation}
\label{optimal}
{\rm Var}_{\rm opt}(u_2) = \frac{4 }{3 d} \,
\frac{3 \ln(1/\epsilon) - 4 + 5 \epsilon - \epsilon^2}{\ln(1/\epsilon)
\left(\ln(1/\epsilon) + 1 + 2 \epsilon\right) - 3 (1 - \epsilon)} \,.
\end{equation}
From Eq.~(\ref{optimal}) we find that in $3d$ ${\rm Var}_{\rm opt}(u_2)
\approx 0.144, 0.096, 0.082$ for $\epsilon = 10^{-3}, 10^{-5},
10^{-6}$, respectively. When $\epsilon \to 0$, ${\rm Var}_{\rm
  opt}(u_2)$ vanishes as
\begin{equation}
\label{asymptotic1}
{\rm Var}_{\rm opt}(u_2) \sim \frac{4}{d} \frac{1}{\ln(1/\epsilon)}\,.
\end{equation}
Therefore, ${\rm  Var}_{\rm opt}(u_2)$  can be made  arbitrarily small
but at  expense of  a progressively higher  resolution.  In  the limit
$\epsilon \to 0$ the  distribution converges to a delta-function.  The
estimators with  $\alpha = 2$ are  the only, in the  family defined by
Eq.~(\ref{u}), that possess this ergodic property.

\section{Diffusion in the presence of a random potential}
\label{sec:disorder}

In this  section we  consider a Brownian  motion in  a one-dimensional
inhomogeneous energy landscape, where  the disorder is correlated over
a finite  length $\xi_c$.   This model gives  a simple  description of
diffusion of a  protein along a DNA sequence,  for instance, where the
particle  interacts with  several  neighboring base  pairs  at a  time
\cite{slutsky}.  The total binding energy of the protein is assumed to
be a  random variable.   When the particle  hops one  neighboring base
further  to  the right  or  to  the left,  its  new  energy is  highly
correlated to the value it had  before the jump.  Slutsky {\it et al.}
\cite{slutsky} modeled this process as a point-like particle diffusing
on a one-dimensional lattice of unit spacing with random site energies
$\{U_i\}$,  whose distribution  is Gaussian  with zero  mean, variance
$\eta^2$      and       is      correlated      in       space      as
$\langle(U_i-U_j)^2\rangle=2\eta^2[1-\exp(-|i-j|/\xi_c)]$,        where
$\xi_c$  is a  correlation length.   At each  time step,  the particle
located  at some  site $i$  jumps to  the left  or to  the  right with
probabilities $p_i \propto  \exp[\beta(U_i-U_{i-1})]$ and $q_i \propto
\exp[\beta(U_i-U_{i+1})]$, respectively, where $p_i+q_i=1$.  Diffusion
is   asymptotically   normal  for   any   disorder  strength   $\eta$.
Nevertheless, the particle  can be trapped in local  energy minima for
long periods of time.  During an extended intermediate time regime, it
is observed  that first passage  properties fluctuate widely  from one
sample to another \cite{slutsky}.

In  ~\cite{boyer1},   we  studied  the   probability  distribution
function $P(u_\alpha)$ with $\alpha=1$  for a Brownian particle moving
in  such a  disordered potential.   There  we found  that $P(u_0)$  is
strongly  affected by  the strength  of the  disorder and  indeed, the
presence of  disorder with short-ranged correlations  tends to broaden
the distribution  of the  measured $D$, as  it presents  an additional
source of fluctuations.

\begin{figure}[!t]
  \centerline{\includegraphics*[width=0.95\textwidth]{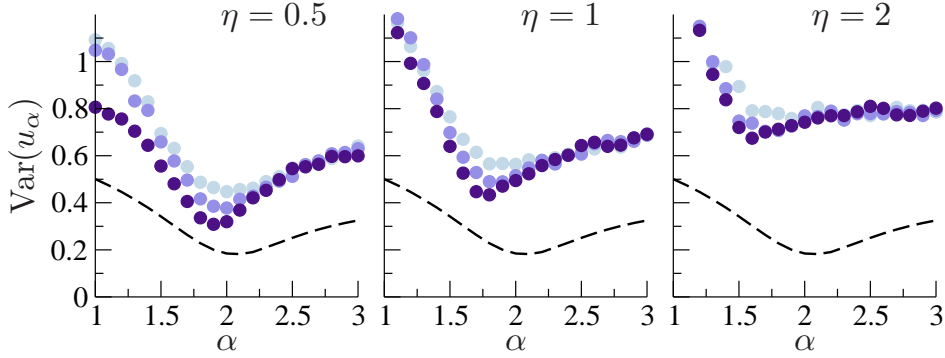}}
  \caption{Variance $\Var{u_\alpha}$ for a Brownian particle moving on
    a short  range disordered potential,  as a function of  $\alpha$ for
    different    values   of    precision:   $\epsilon    =   10^{-4}$
    (${\color{blue1}     \newmoon}$),    $10^{-5}$    (${\color{blue2}
      \newmoon}$)  and $10^{-6}$  (${\color{blue3}  \newmoon}$).  Each
    panel corresponds  to different disorder strengths  as indicated by
    the labels,  and $\beta=1$.  The  dashed curve corresponds  to the
    pure Brownian motion ($\eta=0$) for $\epsilon=10^{-6}$.}
\label{fig4}
\end{figure}
\begin{figure}[!b]
  \centerline{\includegraphics*[width=0.85\textwidth]{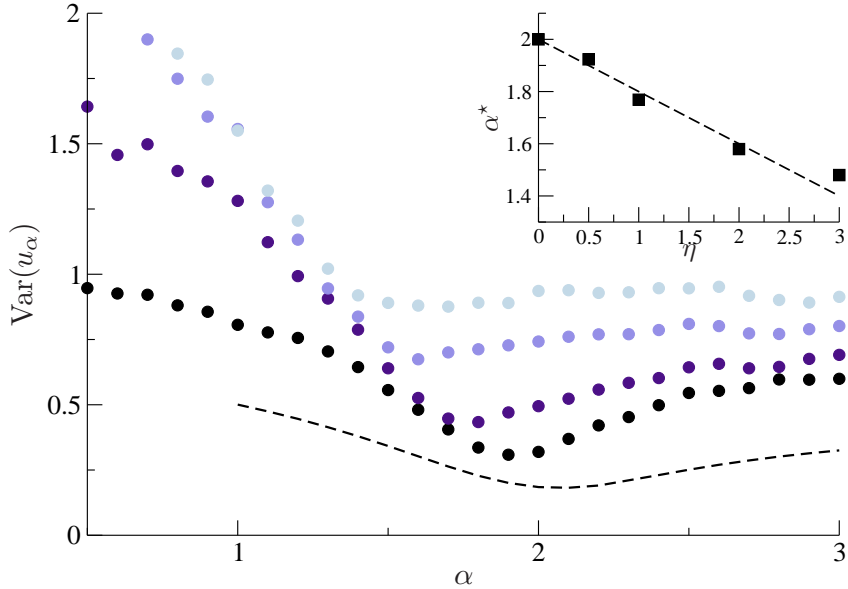}}
  \caption{Variance $\Var{u_\alpha}$ for a Brownian particle moving on
    a   disordered  potential,   as   a  function   of  $\alpha$   for
    $\epsilon=10^{-6}$ and different  values of the disorder strength:
    $\eta  =  3$  (${\color{blue1} \newmoon}$),  $2$  (${\color{blue2}
      \newmoon}$),   $1$   (${\color{blue3}   \newmoon}$)  and   $0.5$
    (${\color{black} \newmoon}$).  The dashed curve corresponds to the
    pure Brownian motion ($\eta=0$).  In the inset, a rough estimation
    of the optimal $\alpha^\star$ as a function of $\eta$.  The dashed
    curve corresponds to $\alpha^\star=2-\eta/5$.}
\label{fig5}
\end{figure}

Here we generalise the study in ~\cite{boyer1} to different values
of  $\alpha$. In  Fig.~\ref{fig4} we  show numerical  results  for the
Variance  $\Var{u_\alpha}$  for  a   Brownian  particle  moving  on  a
disordered  potential  for  different  disorder strengths  $\eta$  and
different precision  $\epsilon$. As expected, for  a fixed $\epsilon$,
the variance  grows in proportion  with the disorder  strength $\eta$.
We do  not expect  that for  these dynamics $u_2$  will be  an ergodic
estimator of  the diffusion  coefficient and as  a matter of  fact, we
cannot  say {\em a  priori}, that  there exists  a value  $\alpha$ for
which  $u_\alpha$  is  ergodic  in  this case.   Nevertheless,  it  is
interesting to note in Fig.~\ref{fig4} that the variance of $u_\alpha$
gets  smaller  around  some  value  $\alpha=\alpha^\star$.   Moreover,
apparently $\alpha^\star\to2$  as $\eta\to0$.  This  behaviour is more
evident   in  Fig.~\ref{fig5}   where  we   show  the   dependence  of
$\Var{u_\alpha}$ on $\alpha$  for a fixed precision $\epsilon=10^{-6}$
and  different   strengths  of   the  disorder  $\eta$.    As  before,
$\Var{u_\alpha}$ seems  to attain  a minimum for  a value  of $\alpha$
which  moves toward  $2$ as  $\eta\to0$.   A rough  estimation of  the
optimal value  $\alpha^\star$ from the numerical data,  shows that the
dependence of  $\alpha^\star$ on $\eta$ is consistent  with the affine
law $\alpha^\star=2-\eta/5$, as shown in the inset of Fig.~\ref{fig5}.

Finally,  to  highlight   the  role  of  the  trajectory-to-trajectory
fluctuations,  we   consider  the  probability  density  function
$P(\omega_{\alpha})$ of the random variable
\begin{equation}
\label{varomega}
\omega_{\alpha} = \frac{u_{\alpha}^{(1)}}{u_{\alpha}^{(1)} + u_{\alpha}^{(2)}} \,,
\end{equation}
where  $u_{\alpha}^{(1)}$  and  $u_{\alpha}^{(2)}$ are  two  identical
independent    random   variables    with   the    same   distribution
$P(u_{\alpha})$.   The  distribution $P(\omega_{\alpha})$,  introduced
recently in  \cite{carlos} (see also \cite{iddo,iddo1,iddo2}),
is  a robust measure  of the  effective broadness  of $P(u_{\alpha})$,
which \textit{probes}  the likelihood of the event  that the diffusion
coefficients drawn  from two different trajectories are  equal to each
other.  This  characteristic property can  be readily obtained  via an
expression \cite{carlos2}
\begin{equation}
 P(\omega_{\alpha}) = \frac{1}{(1 - \omega_{\alpha})^2} \, \int^{\infty}_0 u 
\, du \, P(u) \,
P\left(\frac{\omega_{\alpha}}{1 - \omega_{\alpha}} u\right) \,.
\end{equation}
Hence, $P(\omega_{\alpha})$ is known once we know $P(u_{\alpha})$.  

\begin{figure}[!t]
  \centerline{\includegraphics*[width=0.85\textwidth]{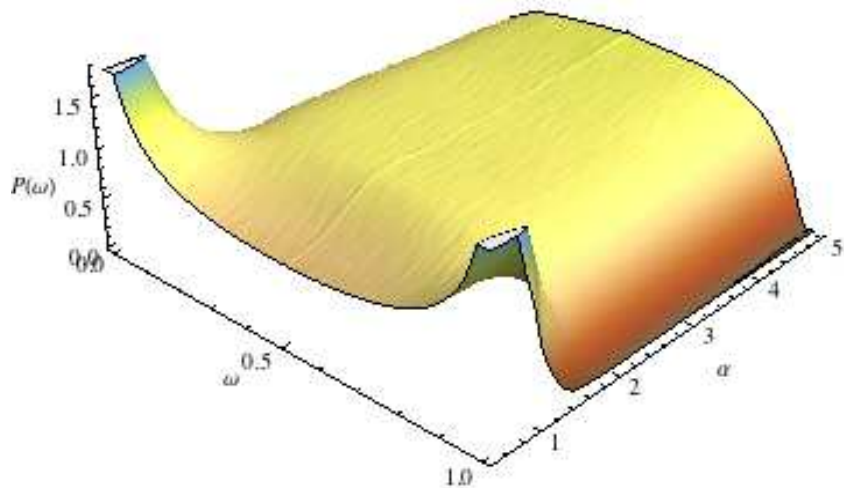}}
  \caption{Distribution  $P(\omega_\alpha)$  as   a  function  of  the
    disorder strength  $\alpha$, for fixed  disorder strength $\eta=3$
    and $\epsilon=10^{-5}$. The colour  ramp increases from red (small
    values) to blue (large values).}
\label{fig6}
\end{figure}

In ~\cite{boyer1},  we observed that for $\eta  \approx 0.8$ the
distribution  $P(\omega_1)$  undergoes  a  continuous  shape  reversal
transition  - from  a unimodal  bell-shaped form  to  a characteristic
bimodal $M$-shape  one with  the minimum at  $\omega_1 = 1/2$  and two
maxima  approaching   $0$  and  $1$  at   larger  disorder  strengths.
Therefore,  indicating  that   for  $\eta  >  0.8$  sample-to-sample
fluctuations  becomes  essential  and  it  is  most  likely  that  the
diffusion coefficients  drawn from two different  trajectories will be
different. Here  we have computed  the distribution $P(\omega_\alpha)$
for different  values of $\alpha$,  $\eta$ and $\epsilon$.   We have
found that  the transition  of $P(\omega_\alpha)$ described  above, is
robust with respect to $\alpha$.  Furthermore, we have also found that
the  same transition  occurs for  fixed strength  of the  disorder and
varying $\alpha$. An example of  this transition is shown as a surface
plot in Fig.~\ref{fig6}. For  $\eta =3$, the transition from bimodal
to  unimodal $P(\omega_\alpha)$  occurs around  $\alpha\approx2$. This
means that in the presence of disorder, among the estimators \eref{u},
those  for  smaller   $\alpha$  present  more  trajectory-to-trajectory
fluctuations than those with larger $\alpha$.

\section{Conclusions}
\label{sec:concl}

To summarize,  we have  discussed different statistical  estimators of
the   diffusion   coefficient   of   a  single   (or   few)   Brownian
single-particle  trajectories. We  have differentiated  between one-time
and two-time estimators. In the  case of the former, we have discussed
a  family of  weighted least-squares  estimators and  showed  that for
standard Brownian  motion there exist a unique  representative of this
family  $u_2$ that possesses  the ergodic  property, meaning  that the
estimated  diffusion coefficient converges  to its  true value  in the
limit of infinite precision $\epsilon\to0$.

Moreover,  we have  sketched the  derivation of  the  full probability
distribution  function   of  $u_\alpha$   in  the  leading   order  in
$\epsilon$. For $\alpha=2$, we have extracted the optimal value of the
variance as a function of  the precision $\epsilon$, and shown that in
the  limit  $\epsilon\to0$,  $\Var{u_e}$  vanishes  in  proportion  of
$1/\ln(T)$.  This means that  for practical purposes the methods based
on  two-time  correlation  functions  can provide  better  estimators,
because the variance of  the corresponding estimator decays faster, as
$1/T$,    even    in    the    presence   of    localization    errors
\cite{berglund,michalet,mb}.

Furthermore, we studied trajectories  of a Brownian particle moving in
a  random  potential  with  short-ranged correlations.  Our  numerical
results    suggest    that    there    exist    an    optimal    value
$\alpha=\alpha^\star$  for   which  the  variance   of  $u_\alpha$  is
minimal.  We  found that  $\alpha^\star$  converges  to $2$  inversely
proportionally to  the strength of the disorder  $\eta$.  Moreover, we
found  that  the  trajectory-to-trajectory  fluctuations  increase  in
proportion to $\eta$ and are inversely proportional to $\alpha$.  This
means that  a robust  estimator of the  diffusion coefficient  for the
random motion  in disordered  potentials is that  for which  the short
time  dynamics is  more efficiently  emphasized. The  question  on the
existence  of an ergodic  estimator of  the diffusion  coefficient for
disordered potentials deserves further investigation.

\begin{acknowledgement}
CMM is partially supported by the European Research Council and the
  Academy of Finland.
\end{acknowledgement}


\begin{thebibliography}{}


\bibitem{perrin}
Perrin J,  C. R. Acad. Sci.   {\bf 146}, (1908) 967;
Ann. Chim. Phys. {\bf 18}, (1909) 5.
%
\bibitem{bra}
Br{\"a}uchle C, Lamb D C and Michaelis J, Eds., 
{\it Single particle tracking and single molecule energy
transfer} (Wiley-VCH, Weinheim 2010).
%
\bibitem{saxton}
Saxton M J and Jacobson K,  Ann. Rev. Biophys. Biomol. Struct. {\bf 26}, (1977)
373.
%
\bibitem{mason}
Mason T G and Weitz D A,  Phys. Rev. Lett. {\bf 74}, (1995) 1250.
%
\bibitem{greenleaf}
Greenleaf W J, Woodside M T and Block S M, 
Annu. Rev. Biophys. Biomol. Struct. {\bf 36}, (2007) 171.
%
\bibitem{weber}
Weber S C, Spakowitz A J and Theriot J A, 
Phys. Rev. Lett. {\bf 104}, (2010) 238102.
%
\bibitem{bronstein}
Bronstein I {\it et al.},  Phys. Rev. Lett. {\bf 103}, (2009) 018102.
%
\bibitem{seisenberger}
Seisenberger G {\it et al.},  Science {\bf 294}, (2001) 1929.
%
\bibitem{weigel}
Weigel A V, Simon B, Tamkun M M and Krapf D,  Proc. Natl. Acad. Sci. USA {\bf 108}, (2011) 6438.
%
\bibitem{golding}
Golding I  and Cox E C, Phys. Rev. Lett. \textbf{96}, (2006) 098102.
%
%
\bibitem{saxton2}
Saxton M J, Biophys. J. {\bf 72}, (1997) 1744.
%
\bibitem{pederson}
Pederson T, Nature Cell Biol. \textbf{2}, (2000) E73-74.
%
\bibitem{rebenshtok}
Rebenshtok A and Barkai E, Phys. Rev. Lett. {\bf 99}, (2007) 210601.
%
\bibitem{ralf}
Jeon J H {\it et al.},  Phys. Rev. Lett. {\bf 106}, (2011) 048103.
%
\bibitem{he}
He Y, Burov S, Metzler R and Barkai E,
Phys. Rev. Lett. {\bf 101}, (2008) 058101.
%
\bibitem{lubelski}
Lubelski A, Sokolov I M and Klafter J,
Phys. Rev. Lett. {\bf 100}, (2008) 250602.
%
\bibitem{barkai} 	
	Schulz J, Barkai E and Metzler R, Ageing effects in single
        particle trajectory averages, Phys. Rev. Lett. {\bf 110} (2013) 020602.



%
\bibitem{carlos}
Mej\'{\i}a-Monasterio C, Oshanin G and Schehr G,
J. Stat. Mech., P06022 (2011).
%
\bibitem{thiago} Mattos T G, Mej\'{\i}a-Monasterio C, Metzler R and Oshanin G, Phys. Rev. E {\bf 86}, (2012) 031143.
%
\bibitem{austin}
Wang Y M, Austin R H and Cox E C,
Phys. Rev. Lett. {\bf 97}, (2006) 048302.
%
\bibitem{goulian}
Goulian M and Simon S M, Biophys. J. {\bf 79}, (2000) 2188.
%
\bibitem{berglund} Berglund A J, Phys. Rev. E {\bf 82}, (2010) 011917.
%
\bibitem{michalet} X.~Michalet,
   Phys. Rev. E {\bf 82}, (2010) 041914; {\bf 83}, (2011) 059904.
%
\bibitem{mb}
X.~Michalet and A.~J.~Berglund, Phys. Rev. E {\bf 85}, (2012)  061916.
%
\bibitem{greb1} Grebenkov D S, Phys. Rev. E {\bf 83}, (2011) 061117.
%
\bibitem{greb2} Grebenkov D S, Phys. Rev. E {\bf 84}, (2011) 031124.
%
\bibitem{greb3} Andreanov A and Grebenkov D S,  J. Stat. Mech. (2012) P07001
%
\bibitem{tej} Tejedor V et al.,  Biophys J {\bf 98}, (2010) 1364.
%
\bibitem{boyer}
Boyer D and Dean D S, J. Phys. A: Math. Gen. {\bf 44}, (2011) 335003.
%
\bibitem{boyer1}
Boyer D, Dean D S, Mej\'{\i}a-Monasterio C and Oshanin G, Phys. Rev. E {\bf 85}, (2012) 031136.
%
\bibitem{qian}
Qian H, Sheetz M P, and Elson E L,, Biophys. J., {\bf 60} (1991) 910.
%
\bibitem{martin}
Martin D, Forstner M, and K\"as J, Biophys. J., {\bf 83} (2002) 2109.
%
\bibitem{savin}
Savin T and Doyle P S, Biophys. J., {\bf 88} (2005) 623.
%
\bibitem{boyer2}
Boyer D, Dean D S, Mej\'{\i}a-Monasterio C and Oshanin G, Phys. Rev. E
{\bf 86} (2012) 060101.
%
\bibitem{boyer3}
Boyer D, Dean D S, Mej\'{\i}a-Monasterio C and Oshanin G, {\tt arXiv:1301.4374}.
%
\bibitem{slutsky}
M. Slutsky, M. Kardar, and L. A. Mirny, Phys. Rev. E {\bf 69}, (2004) 061903.
%
\bibitem{abramowitz}
Abramowitz M and Stegun I R, Eds.,  
{\it Handbook of mathematical functions} (Dover, New York 1972).
%
\bibitem{iddo} Eliazar I, Physica A {\bf 356}, (2005)  207.
%
\bibitem{iddo1} Eliazar I and Sokolov I M, Journal of Physics A: Mathematical and Theoretical {\bf 43}, (2010) 055001.
%
\bibitem{iddo2} Eliazar I and Sokolov I M, Physica A {\bf 391}, (2012) 3043.
%
\bibitem{carlos2}
 Mej\'{\i}a-Monasterio C, Oshanin G and Schehr G,
Phys. Rev. E {\bf 84}, (2011) 035203.
\end{thebibliography}
\end{document}